\documentclass[fleqn,10pt]{wlscirep}
\usepackage[utf8]{inputenc}
\usepackage[T1]{fontenc}
\usepackage{amsmath}
\usepackage{amsfonts}
\usepackage{bbm}
\usepackage{graphicx}
\usepackage{mathtools}
\usepackage{bm}
\usepackage{gensymb}
\usepackage{hyperref}
\usepackage{float}
\usepackage[font=small,labelfont=bf,
   justification=justified,
   format=plain]{caption}
\usepackage{multirow}
\usepackage[super]{nth}
\usepackage{booktabs}

\newcommand{\etal}{\textit{et al}.}
\newcommand{\ie}{\textit{i}.\textit{e}.}
\newcommand{\eg}{\textit{e}.\textit{g}.}

\newcommand{\qts}[1]{``#1''}

\DeclareMathOperator*{\argmin}{arg\,min}
\DeclareMathOperator*{\Tr}{Tr}
\newcommand{\norm}[1]{\left\lVert#1\right\rVert}
\newcommand{\R}{\mathbb{R}}

\title{Anatomically-Informed Deep Learning on Contrast-Enhanced Cardiac MRI for Scar Segmentation and Clinical Feature Extraction}

\author[1,$\dag$]{Haley G. Abramson}
\author[2,$\dag$]{Dan M. Popescu}
\author[3]{Rebecca Yu}
\author[1]{Changxin Lai}
\author[1]{Julie K. Shade}
\author[4]{Katherine C. Wu}
\author[2]{Mauro Maggioni}
\author[1]{Natalia A. Trayanova*}
\affil[$\dag$]{These authors contributed equally to this work}
\affil[1]{Department of Biomedical Engineering, Johns Hopkins University School of Medicine, Baltimore, MD, 21205, USA.}
\affil[2]{Department of Applied Mathematics and Statistics, Johns Hopkins University, Baltimore, MD, 21218, USA.}
\affil[3]{Department of Biomedical Engineering, Johns Hopkins University, Baltimore, MD, 21218, USA.}
\affil[4]{Division of Cardiology, Department of Medicine, Johns Hopkins Hospital, Baltimore, MD, 21205, USA.}
\affil[*]{Corresponding Author}

\begin{abstract}
Visualizing disease-induced scarring and fibrosis in the heart on cardiac magnetic resonance (CMR) imaging with contrast enhancement (LGE) is paramount in characterizing disease progression and quantifying pathophysiological substrates of arrhythmias. However, segmentation and scar/fibrosis identification from LGE-CMR is an intensive manual process prone to large inter-observer variability. Here, we present a novel fully-automated anatomically-informed deep learning solution for left ventricle (LV) and scar/fibrosis segmentation and clinical feature extraction from LGE-CMR. The technology involves three cascading convolutional neural networks that segment myocardium and scar/fibrosis from raw LGE-CMR images and constrain these segmentations within anatomical guidelines, thus facilitating seamless derivation of clinically-significant parameters. In addition to available LGE-CMR images, training used \qts{LGE-like} synthetically enhanced cine scans. Results show excellent agreement with those of trained experts in terms of segmentation (balanced accuracy of $96\%$ and $75\%$ for LV and scar segmentation), clinical features ($2\%$ difference in mean scar-to-LV wall volume fraction), and anatomical fidelity. Our segmentation technology is extendable to other computer vision medical applications and to problems requiring guidelines adherence of predicted outputs.
\end{abstract}
\begin{document}

\flushbottom
\maketitle

\thispagestyle{empty}

\section*{Introduction}
Many cardiac diseases are associated with structural remodeling of the myocardium. In both ischemic and non-ischemic cardiomyopathies, the presence and extent of myocardial fibrosis and scar significantly elevate the risk for lethal heart rhythm disorders and sudden cardiac death (SCD)~\cite{Scott2011, Zipes1998, Roes2009,Kwon2009,Kuruvilla2014_Wu,DiMarco2017_Wu}. Identifying and quantifying the extent of myocardial scar and fibrosis are clinically valuable in assessing arrhythmia propensity in the heart and stratifying patients for SCD risk~\cite{Scott2011,Burg2003,Aljaroudi2013_Wu,Wu2017}.

Cardiac magnetic resonance (CMR) imaging with late gadolinium enhancement (LGE) has unparalleled capability to visualize myocardial scar and fibrosis~\cite{Kim1999}, allowing for direct characterization of the pathophysiological substrate underlying the propensity for arrhythmias and SCD. A large body of research~\cite{Roes2009,disertori2016myocardial,mordi2014lge,cain2014cardiac,marra2014impact,DiMarco2017_Wu,Kuruvilla2014_Wu,Aljaroudi2013_Wu} has demonstrated the utility of scar/fibrosis (\ie, image enhancement) characterization via LGE-CMR to advance our understanding of the mechanisms underlying arrhythmogenesis and in the diagnosis and treatment of ventricular tachyarrhythmias.

Currently, LGE-CMR image analysis requires significant training and expertise. Moreover, it is labor-intensive as it requires slice-by-slice identification and time-consuming contouring of normal myocardium and the distinct regions of enhancement~\cite{Zhuang2016,Zhuang2019}. Both blood pool and scar have high intensities on LGE, posing a challenge for humans and computer programs alike in distinguishing the endocardial border~\cite{Campello2020} and hindering the use of traditional thresholding methods~\cite{Suri2000,Huang2011}. Methods that claim to automatically identify enhancement require manual delineation of the endocardium and epicardium, and variable inter-reader consideration of intermediate signal intensities can increase the variance in scar/fibrosis segmentation~\cite{Klem2017_Wu}. Inter-observer variability is further exacerbated by the fact that low-intensity, viable myocardium is co-localized within the same anatomical structure as high-intensity enhanced tissue. Image segmentation methods such as region growing require individually selected parameters for each scan due to noise differences across MR scanners and imaging centers, in addition to significant computational resources~\cite{Suh1993,atriachallenge}. Semi-automated approaches such as graph cuts~\cite{graphcuts} still demand frequent manual overrides to predicted segmentations~\cite{Lu2012_Wu}. Fully automating segmentation of the left ventricle (LV) with regions of enhancement could dramatically increase the utility of LGE-CMR by improving scar/fibrosis quantification accuracy and reproducibility and reducing image analysis time. Such an advancement can substantially improve the design of clinical trials that use infarct size as an outcome and impact clinical care by more accurately identifying patients at risk for post-infarction outcomes such as heart failure, ventricular arrhythmias, and SCD.

Deep learning (DL) applied to cardiac image segmentation offers the promise of full automation and consistency of output~\cite{Chen2020}. However, most of the available algorithms still require intensive manual interventions, \eg, specifying anatomical landmarks~\cite{Bello2019} or labeling boundary slices of the stack at the apex and base of the heart~\cite{Zheng2018}. The few DL algorithms developed for LGE-CMR myocardial segmentation~\cite{Yue2019,Roth2020,ChenOuyang2019,Campello2020} and the even fewer for LGE-CMR scar/fibrosis segmentation~\cite{Zabihollahy2020, moccia2019development, Fahmy2019_Wu} all suffer from significant limitations. Specifically, these approaches are not robust to varying image acquisition quality (\ie, different scanners and protocols at different centers) or to the varying fibrosis patterns resulting from different heart pathologies, leading to bespoke algorithms which fail to generalize across populations~\cite{Zabihollahy2020,Fahmy2019_Wu}. 
Additionally, the existing small number of LGE-CMR datasets can impair data-intensive DL algorithms, further diminishing segmentation accuracy and scar/fibrosis tissue recognition. Furthermore, because success has been measured by broad study-wide averages that often elicit poor-performing outliers~\cite{khened2019fully}, anatomical fidelity of computer-generated cardiac segmentations is often compromised~\cite{Zabihollahy2020}, resulting in inability to extract features of significance to clinical decision-making. Therefore, a data-efficient, automated algorithm for anatomically-accurate LGE-CMR segmentation of normal myocardium and scar/fibrosis remains an unmet clinical need.

Here we describe an anatomically-informed deep-learning (DL) approach to myocardial and scar/fibrosis segmentation and clinical feature extraction from LGE-CMR images. We term our technology Anatomical Convolutional Segmentation Network, ACSNet. The technology enables clinical use by ensuring anatomical accuracy and complete automation. It is robust to inputs from different CMR modalities and imaging centers, and scar/fibrosis distributions arising from different heart pathologies. Unlike existing LGE-CMR segmentation technologies, ours accurately quantifies scar and calculates LV volume, outperforming inter-expert results without compromising the anatomical integrity of the segmentation. By eliminating manual interventions, this approach holds the potential to make CMR more accessible and less labor-intensive. Furthermore, ACSNet's approach for clinical feature extraction, which satisfies highly complex geometric constraints without stunting the learning process, has the potential for broad applicability in computer vision beyond cardiology, and even outside of medicine.

\section*{Anatomically-Informed Deep Learning Model}
ACSNet uses three inter-connected DL sub-networks that progressively refine the LV myocardial and enhancement segmentation from the LGE-CMR images. The input to the first DL sub-network is the result of an innovative data augmentation process, which overcomes the scarcity of public LGE-CMR scans by transforming more readily available CMR cine scans into \qts{LGE-like} images. To accomplish this, we developed a low-cost cine-to-LGE conversion algorithm that adds pseudo-enhancement to non-enhanced cine. The process tripled the available dataset to $400$ patient scans, amounting to $2,484$ images~\cite{ACDC,Sunnybrook,Wu2020Prose}, acquired by a wide variety of imaging protocols in heterogeneous patient populations.

\begin{figure}[ht]
\centering
\includegraphics[width=\linewidth]{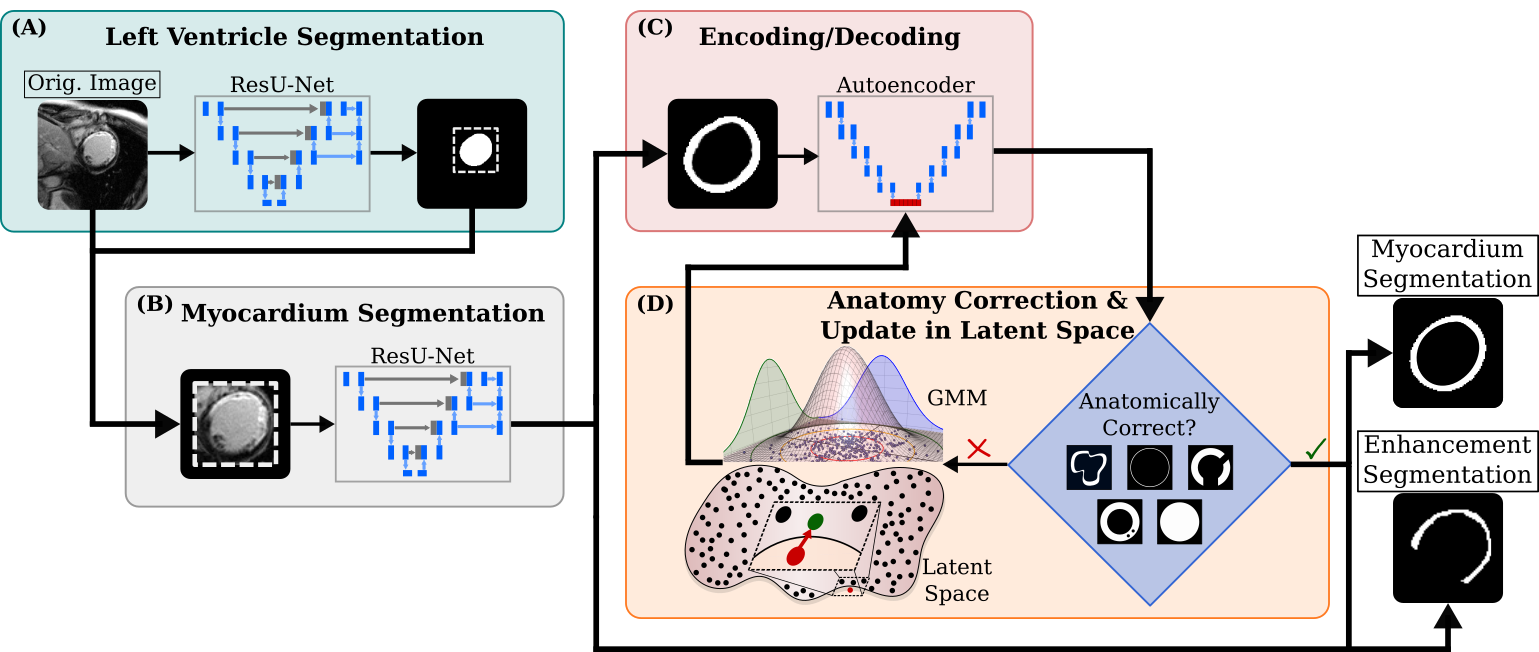}
\caption{ \textbf{ACSNet architecture consisting of three inter-connected DL sub-networks}. \textbf{(A)} The first residual U-Net (ResU-Net) is used to identify and crop around the left ventricle (LV). \textbf{(B)} The second network uses the tightly cropped image from (A) and the LV segmentation to further segment the LV into viable and enhanced myocardium. \textbf{(C)} The third network is a convolutional autoencoder trained to encode (compress) and decode myocardial segmentation masks. \textbf{(D)} Segmentations from the training set are encoded using the third network to form a latent space. The space is modeled as a Gaussian Mixture Model (GMM) and conditional re-sampling is performed to populate the space with anatomically correct samples (black dots). Predicted segmentations are encoded and the nearest neighbors algorithm is used to return a perturbed, anatomically correct version (green dot) of the original (red dot). GMM image adapted from source~\cite{GMM_figure}. Additional details are presented in Methods.}
\label{fig:network_arch}
\end{figure}

The first network in ACSNet handles the overwhelmingly high ratio of background pixels to myocardium by identifying the LV and cropping each image tightly around it (fig.~\ref{fig:network_arch}A). The second network then uses the tightly cropped image to differentiate blood pool, viable myocardium, and regions of enhancement, returning segmentations for both tissue types (fig.~\ref{fig:network_arch}B). The third network is a convolutional autoencoder which refines the shape of the myocardium and scar/fibrosis in each slice and in the total LV volume (fig.~\ref{fig:network_arch}C). 

Correct training segmentations are encoded into the autoencoder latent space; we model their distribution as a Gaussian mixture model (GMM). We then populate the latent space by sampling from the GMM but rejecting samples that, when decoded into their image representation, violate anatomical constraints (fig.~\ref{fig:network_arch}D). Upon predicting a new segmentation, we encode the input image and perturb the point in the encoding space in the direction of the closest point in the population sampled from the constrained GMM until anatomical constraints are fulfilled by the point's decoded image representation, the latter being the final post-processed output. As a result, predicted segmentations by ACSNet pass complex whole-ventricle anatomical requirements such as circularity and minimum thickness (see Methods), which not only ensure anatomical accuracy, but also allow for proper handling of ambiguous regions (\eg, apex and base), where observer ground truth variability is high due primarily to imaging artifacts.

\section*{Application-Driven Evaluation of the Technology}

\begin{figure}[ht]
\centering
\includegraphics[width=0.75\linewidth]{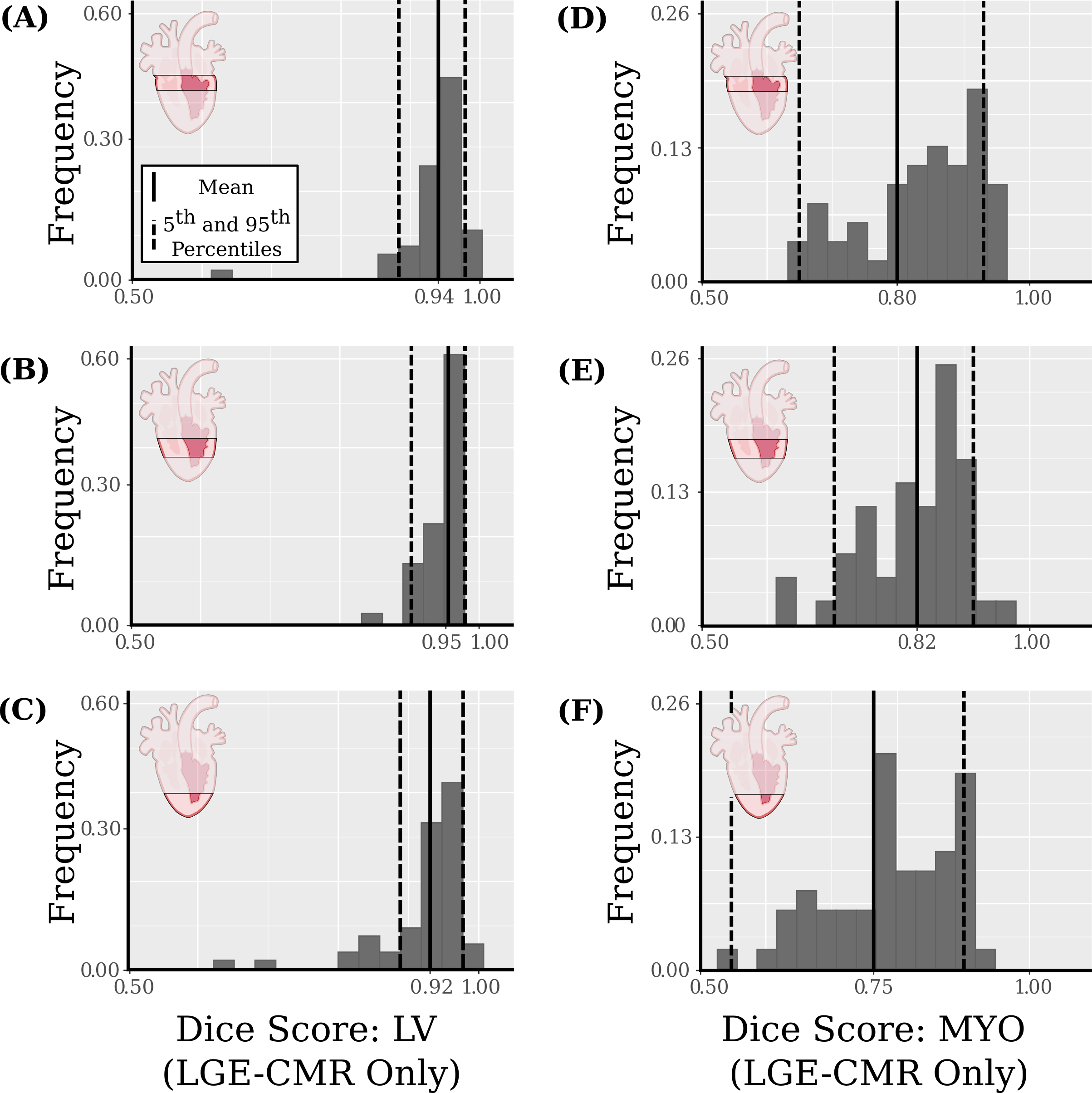}
\caption{ \textbf{Left Ventricle and Myocardium Segmentation Results by Region}. Histograms of per-slice Dice scores are shown for three regions of the heart (rows: basal, middle, and apical). \textbf{(A}-\textbf{C)} Dice scores are shown for the left ventricle (LV) and \textbf{(D}-\textbf{F)} myocardium (MYO). The averages (solid vertical lines) for each region are $0.94$ \textbf{(A)} and $0.80$ \textbf{(D)} for basal slices, $0.95$ \textbf{(B)} and $0.82$ \textbf{(E)} for middle slices, and $0.92$ \textbf{(C)} and $0.75$ \textbf{(F)} for apical slices. A typical stack consists of 12 slices roughly equally distributed in the three regions.}
\label{fig:dice_results_fig}
\end{figure}

\begin{figure}[ht]
\centering
\includegraphics[width=.98\linewidth]{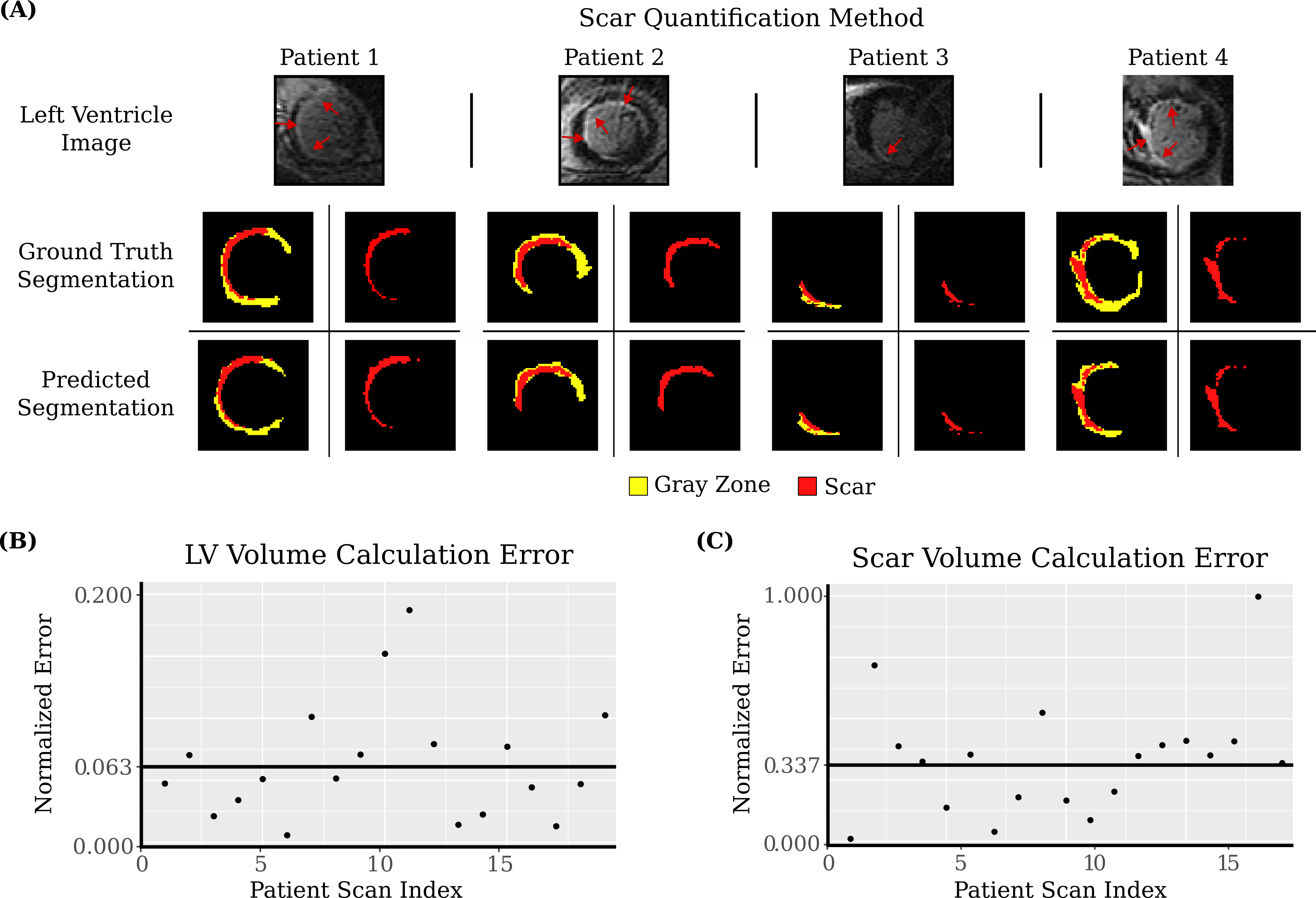}
\caption{\textbf{Scar and LV Volume Calculation Results}. \textbf{(A)} Segmentations of enhancement regions from the left ventricle (LV) segmentation network represent gray zone (GZ, yellow) and scar (red). We quantify the scar by thresholding the regions of enhancement at the full width-half maximum (FWHM) intensity value of the masked image. We illustrate the original scan (first row), the ground truth scar and GZ segmentations (middle row), and the predicted segmentations (bottom row). Patients 1-3 are representative examples of scar and GZ segmentations. Patient 4 is an outlier for which GZ segmentation has low accuracy; this, however, does not affect the scar segmentation. \textbf{(B)} LV volume was calculated as the sum of myocardium and blood pool volumes from both predicted and manual segmentations. Error for LV and scar volumes is calculated as the absolute error normalized by each respective volume. The normalized mean absolute error (MAE) of LV volumes was $6.3\%$ (solid black line). Each point represents the error in LV volume of a single segmented patient scan. \textbf{(C)} All scar volume errors are plotted. The normalized MAE is $33.7\%$ (solid black line).  }
\label{fig:scar_results_fig}
\end{figure}

In training ACSNet, we used LGE-CMR scans from patients with ischemic cardiomyopathy as well as cine scans from numerous cohorts that we synthetically enhanced to appear \qts{LGE-like}. ACSNet's performance was evaluated by testing only on out-of-sample LGE-CMR scans ($269$ images, accounting for $25\%$ of the LGE cohort) that have the true capability of visualizing scar/fibrosis, while both cine-converted \qts{LGE-like} and LGE-CMR images were used during training ($2,484$ images). The LV segmentation network identified blood pool, viable, and enhanced tissue, the latter consisting of core scar and intermediate intensity peri-infarct or gray zone (GZ) --- both demonstrated to have distinct contributions to arrhythmogenesis~\cite{SchmidtWu2007}. LV segmentations resulted in Dice scores of $0.96$ and $0.93$ in training (on both LGE-CMR and cine) and testing (on LGE-CMR), respectively. The second sub-network identified the LV myocardium with a Dice score of $0.91$ in training and tested on LGE-CMR images with a Dice score of $0.79$ for the LV wall (myocardium) segmentation. With anatomical post-processing by the third sub-network, the myocardial  segmentation maintained a Dice score of $0.79\pm0.02$, with $93\%$ balanced accuracy (fig.~\ref{fig:dice_results_fig}). The Hausdorff distance (HD), highly sensitive to even very small differences, measured from ACSNet predictions ($6.72\pm0.58$mm) was lower than that of previously recorded LGE-CMR segmentation methods (see table~\ref{tab:seg_comparison}), indicating increased similarity between predicted and ground truth segmentations. Thus, ACSNet improved upon both the inter-observer Dice score of $0.76$ as well as the inter-observer HD ($10.6\pm4.65$ endocardial HD and $12.5\pm5.38$ epicardial HD) recorded in the Multi-Sequence Cardiac MRI Segmentation Challenge~\cite{Zhuang2016,Zhuang2019}. While comparable Dice scores between ACSNet and existing DL applications indicate a level playing field (see table~\ref{tab:seg_comparison}), the power of ACSNet lies in its consistency of segmentation across all LV regions and its guarantee of anatomical accuracy, which has not been previously achieved for LGE-CMR. Our results demonstrate no statistically significant difference between features calculated using automatic vs. manual segmentations. LV volume calculations result in a relative (normalized by the LV volume) mean absolute error (MAE) of $6.3$\% (fig.~\ref{fig:scar_results_fig}\textbf{(B)}, table~\ref{tab:results_slice}). Scar volume calculation has a relative MAE (normalized by scar volume) of $33.7$\% (fig.~\ref{fig:scar_results_fig}\textbf{(C)}, table~\ref{tab:results_vol}). 

\begin{table}[ht]
\centering
\begin{tabular}{@{}llrrrr@{}}
\toprule
\multicolumn{2}{c}{\multirow{2}{*}{Measure}} & \multicolumn{3}{c}{Location} & \multicolumn{1}{c}{\multirow{2}{*}{Total}} \\ \cmidrule(lr){3-5}
\multicolumn{2}{c}{} & \multicolumn{1}{c}{Apex} & \multicolumn{1}{c}{Middle} & \multicolumn{1}{c}{Base} & \multicolumn{1}{c}{} \\ \cmidrule(r){1-2} \cmidrule(l){6-6} 
LV &  & \multicolumn{1}{l}{} & \multicolumn{1}{l}{} & \multicolumn{1}{l}{} & \multicolumn{1}{l}{} \\
 & BA (\%) & $96.3\pm0.4$ & $96.3\pm0.4$ & $96.2\pm0.5$ & $96.3\pm0.2$ \\
 & Dice & $0.92\pm0.02$ & $0.95\pm0.01$ & $0.94\pm0.01$ & $0.93\pm0.01$ \\
 & HD (mm) & $6.9\pm1.6$ & $6.1\pm0.8$ & $6.6\pm1.5$ & $6.5\pm0.8$ \\
MYO &  &  &  &  &  \\
 & BA (\%) & $93.3\pm1.1$ & $93.1\pm1.2$ & $92.9\pm1.4$ & $93.1\pm.7$ \\
 & Dice & $0.75\pm0.04$ & $0.82\pm0.02$ & $0.80\pm0.04$ & $0.79\pm0.02$ \\
 & HD (mm) & $6.4\pm0.7$ & $6.6\pm0.7$ & $7.2\pm1.5$ & $6.7\pm0.6$ \\
Enhancement Region &  &  &  &  &  \\
 & BA (\%) & $69.9\pm2.3$ & $69.8\pm2.6$ & $70.4\pm2.6$ & $70.0\pm1.4$ \\
 & Dice & $0.51\pm0.06$ & $0.48\pm0.07$ & $0.59\pm0.09$ & $0.51\pm0.04$ \\
 & HD (mm) & $16.8\pm3.4$ & $24.0\pm6.6$ & $19.8\pm8.7$ & $19.9\pm3.3$ \\
Core Scar Region &  &  &  &  &  \\
 & BA (\%) & $74.9\pm2.8$ & $74.3\pm3.1$ & $75.5\pm3.3$ & $74.9\pm1.8$ \\
 & Dice & $0.57\pm0.08$ & $0.52\pm0.09$ & $0.63\pm0.11$ & $0.57\pm0.05$ \\
 & HD (mm) & $14.9\pm3.7$ & $24.4\pm6.7$ & $18.1\pm8.9$ & $18.9\pm3.5$ \\
\multicolumn{1}{l}{} & \multicolumn{1}{l}{} & \multicolumn{1}{l}{} & \multicolumn{1}{l}{} \\ \bottomrule
\end{tabular}
\caption{\textbf{ACSNet LGE-CMR Segmentation Performance.} Balanced accuracy (BA), Dice coefficient (Dice), and Hausdorff distance (HD) are shown for 4 regions of interest segmented by ACSNet: whole left ventricle (LV), myocardial tissue (MYO), area of enhancement (Enhancement Region), and scar tissue (Core Scar Region). BA is expressed in percentage terms, Dice is adimensional, and HD is in millimeters. All numbers are averages $\pm$ $95\%$ confidence interval size over apex/middle/base/total slices of all patients in the test set.}
\label{tab:results_slice}
\end{table}

\begin{table}[ht]
\centering
\begin{tabular}{@{}llrrrr@{}}
\toprule
\multicolumn{2}{c}{\multirow{2}{*}{Feature}} & \multicolumn{3}{c}{LV Volume Tertile} & \multicolumn{1}{c}{\multirow{2}{*}{Total}} \\ \cmidrule(lr){3-5}
\multicolumn{2}{c}{} & \multicolumn{1}{c}{Lower} & \multicolumn{1}{c}{Middle} & \multicolumn{1}{c}{Upper} & \multicolumn{1}{c}{} \\ \cmidrule(r){1-2} \cmidrule(l){6-6} 
LV &  & \multicolumn{1}{l}{} & \multicolumn{1}{l}{} & \multicolumn{1}{l}{} & \multicolumn{1}{l}{} \\
& GT (cc) & 226 (186 -- 259) & 307 (280 -- 327) & 405 (334 -- 573) & 312 (186 -- 573)\\
& Pred (cc) & 237 (193 -- 273) & 312 (279 -- 339) & 424 (342 -- 614) & 323 (193 -- 614)\\
& Norm. MAE (\%) & 10.3 (4.8 -- 18.8) & 4.5 (0.6 -- 8.0) & 4.4 (1.3 -- 10.3) & 6.3 (0.6 -- 18.8)\\
MYO &  &  &  &  &  \\
& GT (cc) & 121 (85 -- 159) & 171 (110 -- 215) & 180 (114 -- 274) & 158 (85 -- 274)\\
& Pred (cc) & 144 (109 -- 186) & 187 (144 -- 226) & 217 (159 -- 351) & 183 (109 -- 351)\\
& Norm. MAE (\%) & 24.1 (12.6 -- 41.1) & 12.7 (1.8 -- 30.8) & 24.7 (2.2 -- 53.7) & 20.1 (1.8 -- 53.7)\\
 Enhancement Region &  &  &  &  &  \\
 & GT (cc) & 27 (15 -- 46) & 24 (3 -- 39) & 30 (7 -- 47) & 27 (3 -- 47)\\
& Pred (cc) & 21 (7 -- 43) & 19 (0 -- 33) & 26 (3 -- 38) & 22 (0 -- 43)\\
& Norm. MAE (\%) & 26.6 (5.9 -- 53.4) & 31.1 (3.8 -- 100.0) & 45.4 (15.9 -- 87.5) & 34.2 (3.8 -- 100.0)\\
Core Scar Region &  &  &  &  &  \\
 & GT (cc) & 13 (6 -- 21) & 11 (1 -- 19) & 17 (4 -- 30) & 13 (1 -- 30)\\
& Pred (cc) & 11 (5 -- 20) & 9 (0 -- 17) & 14 (2 -- 22) & 12 (0 -- 22)\\
& Norm. MAE (\%) & 14.4 (1.2 -- 41.2) & 42.5 (18.2 -- 100.0) & 42.7 (32.1 -- 72.0) & 33.7 (1.2 -- 100.0)\\
\multicolumn{1}{l}{} & \multicolumn{1}{l}{} & \multicolumn{1}{l}{} & \multicolumn{1}{l}{} \\ \bottomrule
\end{tabular}
\caption{\textbf{ACSNet LGE-CMR Clinical Feature Performance.} Ground truth (GT) and predicted (Pred.) volumes and mean absolute error normalized by GT volume (Norm. MAE), together with ranges (parentheses) are shown for 4 regions of interest segmented by ACSNet: whole left ventricle (LV), myocardial tissue (MYO), area of enhancement (Enhancement Region), and scar tissue (Core Scar Region). GT and Pred. are expressed in cubic centimeters and Norm. MAE in percentage terms. Numbers represent averages across all patients in the test set (Total) and patients grouped by GT LV volume tertile (Lower/Middle/Upper).}
\label{tab:results_vol}
\end{table}

\section*{Discussion}
In this study we present a clinical-application-driven approach for automatic segmentation, with high anatomical accuracy, of normal myocardium and scar/fibrosis measured from LGE-CMR images and for extracting anatomical features that are essential for disease diagnosis and clinical decision-making. Our ACSNet is a cascade of three neural networks that progressively refine LV myocardial and enhanced tissue segmentations and constrain results within anatomical guidelines for both individual slices and the entire LV volume. These constraints prevent poorly segmented outliers that can arise particularly at the apex and base. Additionally, the diverse dataset used to train ACSNet enhances the network's robustness, expanding its ability to be successfully used in unseen patient populations. Our technology can be seamlessly employed to extract salient anatomical features, potentially enhancing and increasing accessibility of the diagnostic utility of LGE-CMR.

There are innovations at every step in our technology that help boost ACSNet's performance. During pre-processing, we augment our training dataset by generating \qts{LGE-like} images from cine, improving performance on the LGE-CMR test set. The complex learning process is divided into three sub-networks, each having distinct tasks: the first reduces class imbalance between the region of interest and background, the second delineates endocardium and epicardium, and the third  ensures anatomical correctness for both slices and volumes. The third neural sub-network expands on previous work~\cite{Painchaud2019}, originally developed for cine segmentation, by incorporating multiple highly complex 2-D and 3-D geometrical constraints, implementing a smaller non-variational network that does not require knowledge of slice location, and simplifying the structure and modeling of the latent space.

Limitations in previously published cine-based segmentation approaches, such as their requirements for additional types of ground-truth labels, poor performance in the more challenging apical and basal regions, and anatomically inaccurate results, prevent their clinical adoption in LGE-CMR segmentation. For example, the method of Zheng~\etal~\cite{Zheng2018} requires a pre-processing step to discard apical and basal slices and a manual curation of \qts{difficult cases}. Similarly, Bello~\etal~\cite{Bello2019} rely on ground-truth anatomical landmark localization. Without such additional annotations in the training data, there is no control of anatomically inconsistent outliers. Other recent methods have proposed post-processing steps to improve the anatomical plausibility of myocardial segmentations from cine images~\cite{Painchaud2019, Larrazabal2020}. Although these algorithms smooth out resulting segmentations, they use generic techniques unable to capture nuances of  heart anatomy~\cite{Larrazabal2020}; they require an already highly accurate segmentation as input to function well~\cite{Painchaud2019}; or they do not incorporate 3-D constraints~\cite{Painchaud2019}. Furthermore, effective LGE-CMR DL segmentation methods are limited for the myocardium and virtually nonexistent for scar/fibrosis. Approaches such as Campello~\etal\cite{Campello2020} --- which use a CycleGAN~\cite{CycleGAN2017} to transform cine into \qts{LGE-like} images --- partially address the LGE-CMR data scarcity to obtain accurate myocardial segmentations, but, in the process, these approaches lose the enhancement features, the salient aspect of LGE-CMR. In our attempt to implement CycleGAN for the cine-to-LGE conversion, we obtained less than $1\%$ Dice coefficient improvement in LV myocardial segmentation. Very recently, Zabihollahy~\etal~\cite{Zabihollahy2020} have attempted myocardial and scar/fibrosis segmentation on 3-D LGE-CMR, but, despite the benefit of a 10-fold increase in the number of short-axis slices per patient, their results show several artifacts, such as disjoint pieces of myocardium. The method of Fahmy~\etal~\cite{Fahmy2019_Wu} exclusively studied 3-D images of patients with hypertrophic cardiomyopathy with unknown generalizability to other cardiomyopathies. Moccia~\etal~\cite{moccia2019development} required manually segmented ground truth myocardium as an additional network input to predict enhancement segmentations and used only 30 patients, all from a single center, split across training and testing. ACSNet overcomes these limitations by providing an efficient DL segmentation algorithm using the standard-of-care widely used 2-D LGE-CMR scans to produce consistent, anatomically accurate results. 

Despite the complexity of LGE-CMR images, we achieved a high Dice score when testing on them. We improved on inter-observer scores\cite{Zhuang2016,Zhuang2019} and quantified scar with low error ($2\%$ difference in mean scar-to-myocardium volume fraction), while maintaining a high degree of anatomical accuracy. Generalization across patient cohorts, MR scanners (\eg, Siemens$^{\text{\textregistered}}$, Philips$^{\text{\textregistered}}$, and General Electric$^{\text{\textregistered}}$), and health centers is often a concern when developing DL algorithms. We mitigate this challenge by training with numerous datasets~\cite{ACDC, Sunnybrook, Wu2020Prose} acquired with different imaging protocols. An additional obstacle associated with image segmentation is training well when inter-observer variability is high, \ie, when ground truth data is noisy. Rather than merely replicating the noisy ground truth, our network focuses on generating reliable segmentations, which we demonstrate through high performance on both Dice score (fig.~\ref{fig:dice_results_fig}, table~\ref{tab:results_slice}) and clinical feature extraction (fig.~\ref{fig:scar_results_fig}, table~\ref{tab:results_vol}). Even outlier patients with poor performing GZ segmentations (see fig.~\ref{fig:scar_results_fig}\textbf{(A)}, Patient 4) maintain low error in scar prediction upon thresholding at the full width-half maximum intensity value. We avoid potential overfitting when building the network by not performing a broad hyperparameter sweep. Exploration of different hyperparameter implementations demonstrated similar Dice score results, thereby indicating the robustness of the core ACSNet model.

Incorporating complex constraints into a DL algorithm may be broadly applied to many computer vision problems. Convolutional neural networks face challenges in learning non-differentiable functions such as ACSNet's anatomical checks; they face similar challenges in, for example, inverse problems or 3-D scene rendering. In applications beyond medicine such as programming an autonomous vehicle to identify pedestrians in a street scene or modeling planets as imperfect spheres, our method has potential to be utilized in any problem demanding specific guidelines from predicted outputs.

\section*{Methods}
ACSNet uses a 3-stage adaptive approach to segment viable myocardium, enhanced myocardium (gray zone and scar), and blood pool, namely: (1) one network identifies the LV used to null out signal in remote (non-LV) regions; (2) a refinement network differentiates between tissue types; and (3) a post-processing stage updates the predictions under anatomical constraints. Novel pre-processing steps increase model performance on LGE-CMR images, and additional post-processing volumetric guidelines ensure anatomically-correct whole-ventricle segmentations.

\subsection*{Data Source and Processing}
ACSNet was developed using LGE-CMR scans from 155 patients with ischemic cardiomyopathy, amounting to $1,124$ images, part of the LV Structural Predictors of Sudden Cardiac Death study (NCT01076660). We reserved $25\%$ of the LGE-CMR data for testing purposes. We developed an innovative data augmentation process that supplemented our dataset with $1,360$ end diastole (ED) images from 245 cine scans from the MICCAI ACDC\cite{ACDC} and Sunnybrook datasets~\cite{Sunnybrook}. All LGE-CMR and cine scans included ground truth myocardial segmentations, and two-thirds of the LGE-CMR scans additionally provided ground truth scar/fibrosis segmentations (for more details see the Training subsection below).

\begin{figure}[ht]
\centering
\includegraphics[width=\textwidth]{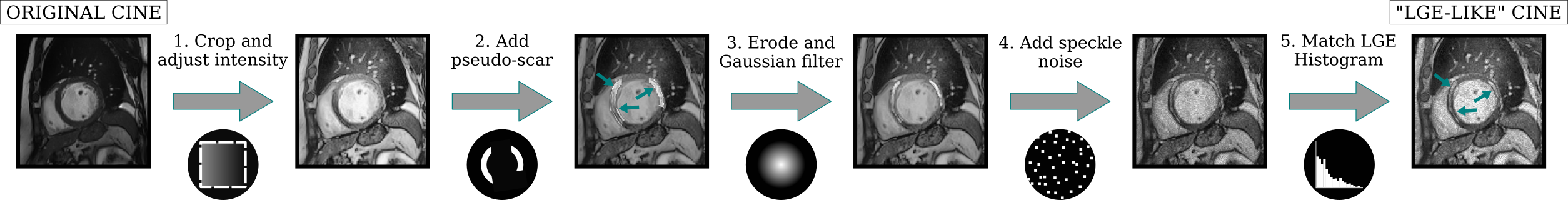}
\caption{\textbf{Conversion Process to \qts{LGE-like} Cine Images}. (1) The original cine image is cropped/padded to a square and contrast-limited adaptive histogram equalization (CLAHE) is applied. (2) Cine images are further transformed by first generating a pseudo-enhancement (\qts{LGE-like} enhanced myocardium) mask. (3) We apply random erosion of the pseudo-scar mask and Gaussian filter to realistically blur the edges. (4) Next, speckle noise is added to the image to resemble LGE noise. (5) Finally, we match the histogram to that of LGE-CMR scan sampled randomly.}
\label{fig:data_aug_fig}
\end{figure}

Ground truth segmentations for cine images are more readily available than LGE, which we take advantage of by augmenting our dataset using \qts{LGE-like} cine images. To perform this style transfer, we generate pseudo-enhancement, or \qts{LGE-like} enhanced myocardium, as follows: first, we randomly sample from the known area of myocardium in a cine image. Then, we blur this area using a Gaussian filter to smooth out the edges, overlay the area as a bright enhancement mask onto the original cine, which contains dark myocardium just as in LGE-CMR, and add speckle noise. Finally, for each cine scan, we randomly sample an LGE scan from the training LGE-CMR cohort, and match the former's intensity histogram to the latter's. (fig.~\ref{fig:data_aug_fig}). 

All image scans were pre-processed and stored in a universally accessible format to account for the large amounts of data required by deep learning algorithms and numerous medical image file types. Although ACSNet takes as input 2-D slices, our file type retained all relevant scan information of the 3-D stack for ease in pre- and post-processing. Slices were ordered apex to base (in increasing vertical direction) with slice location, image intensities, resolution, and patient orientation information stored. Slices with no ground truth segmentation were excluded. We used comprehensive volumetric checks (described below) to trim extraneous slices below the apex and above the base. 

The next step in data pre-processing was standardizing the images by applying rotations and brightness adjustments. Each scan is rotated in increments of $90\degree$ --- thus avoiding rotation artifacts --- to approximately standardize orientation across patient scans. This step aids the training of the neural network, as it eliminates the need for rotational invariant features by ensuring that the LV is
in a similar location on each image, and that the right ventricle and the LV will have the same orientation relative to each other. If tags such as  "WindowCenter", "WindowLength", "RescaleSlope", and "RescaleIntercept" are included on the original file format (\eg, DICOM), they are used to transform between raw signal intensities and display value intensities, enhancing contrast and brightness. 

\subsection*{Model}

\subsubsection*{Stage 1: Left Ventricle Segmentation Network}
The objective of the LV segmentation network (fig.~\ref{fig:network_arch}A) is to reduce the background and clip the image to a square around the LV, reducing the non-myocardium features that the second network could learn. LV segmentation network inputs are modified by applying contrast-limited adaptive histogram equalization (CLAHE)~\cite{Pizer1987} to increase the contrast between myocardium and blood pool. Then, these input images are cropped or padded to a square of size $192\times192$ pixels to eliminate potential aspect ratio distortion. Finally, we scale the intensity values of each image slice between 0 and 255 based on the maximum and minimum intensity values of its cropped pixel array, excluding the added black space.

The LV segmentation network is a U-Net with residuals (ResU-Net) of depth four~\cite{UNet2015,Zheng2018}. During the downsampling process, each of the four depth levels consists of 2 repetitions of a block of a $3\times3$ 2-D convolution, followed by a ReLU activation and batch normalization. After the second set of these three layers, each block ends with a $2\times2$ max pooling layer and $20$\% dropout. A transition layer bridges the downsampling and upsampling branches, each containing similar blocks. However, upsampling is performed with a $2\times2$ convolution (see fig.~\ref{fig:resunet_model}). Predictions are automatically cleaned up to identify the single component most likely to represent both the LV myocardium and the blood pool by performing slice- and volume-level checks (see code for further details upon publication). By cropping inputs to the second network around the LV identified in the first network, we substantially decrease class imbalance between myocardium and signal from the rest of the scan.

\subsubsection*{Stage 2: Myocardium Segmentation Network}
The cropped area containing the LV segmentation as defined by the first network is centered on a $128\times128$ pixel image. Intensities are then scaled based on the histogram of the scan's entire volume rather than individually, as in the LV segmentation network, to ensure consistency in the intensity contrast of enhanced and non-enhanced myocardium. Notably, we do not perform CLAHE. Instead, we normalize the image intensities using the median intensity of the blood pool across the entire volume. Specifically, the following functions are applied sequentially to each input image component-wise:
\begin{align*}
    I &\mapsto \frac{I}{2 m_{I_\mathcal{D}}} \\ 
    I &\mapsto 255 \times \frac{I - \min_{\mathcal{D}}I}{\max_{\mathcal{D}}I - \min_{\mathcal{D}}I} \\ 
    I &\mapsto \min\lbrace255, \max\lbrace0, I\rbrace\rbrace 
\end{align*}
where $I$ is the image intensity, $\mathcal{D}$ is the effective region following cropping by the LV segmentation network, and  $m_{I_\mathcal{D}}$ is the median signal intensity over $\mathcal{D}$.

\begin{figure}[ht]
\centering
\includegraphics[height=3.5in]{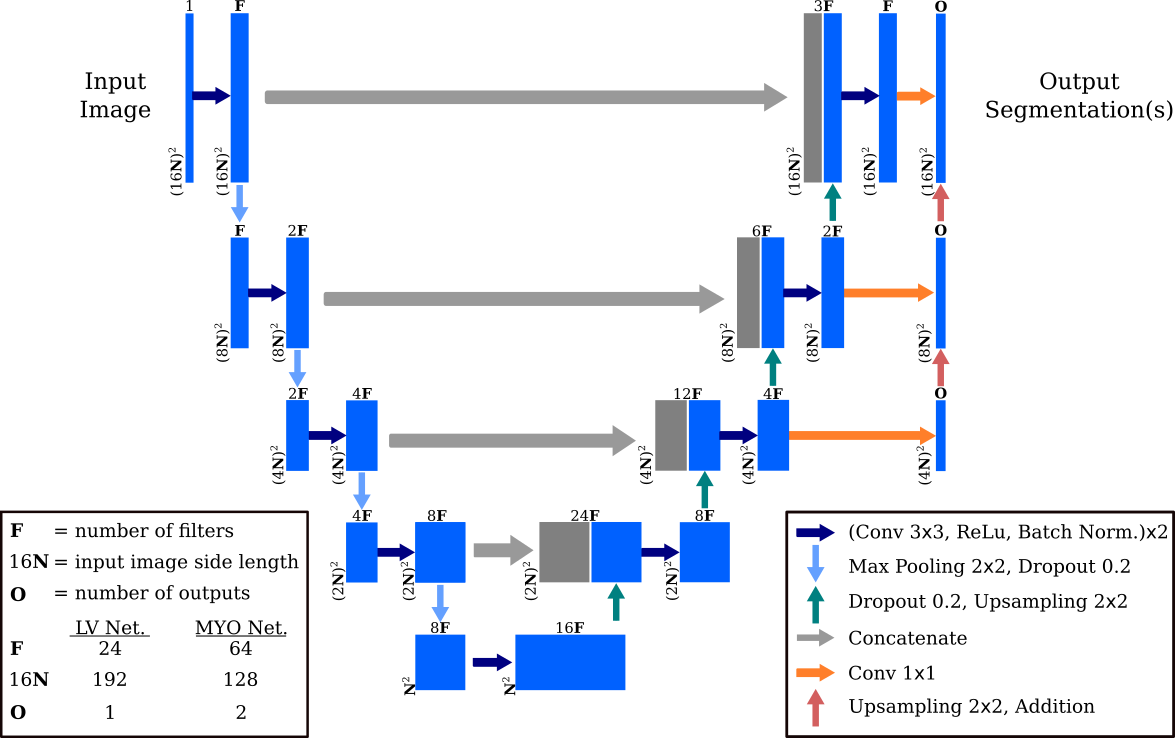}
\caption{ \textbf{Left Ventricle and Myocardium Segmentation Network Architecture}. The left ventricle (LV) network identifies the main region of interest. The second network segments the myocardium (MYO) by differentiating between viable and non-viable tissue represented by each of the two outputs. The networks differ by the number of filters, input image size, and number of outputs as indicated. Both networks are U-Nets with residuals. }
\label{fig:resunet_model}
\end{figure}

Similarly to the LV segmentation network, we implement a ResU-Net structure with some modifications to identify the myocardium (fig.~\ref{fig:network_arch}B). The myocardium segmentation network differs from the LV segmentation network in that it uses twice the number of filters at each of the 4 depth levels due to the smaller image inputs (see fig.~\ref{fig:resunet_model}). It outputs two masks, one representing the entire myocardium and the other identifying only the enhanced tissue. 

\subsubsection*{Stage 3: Anatomical Autoencoder Post-Processing}
\label{anatae_section}
\begin{figure}[ht]
\centering
\includegraphics[width=7in]{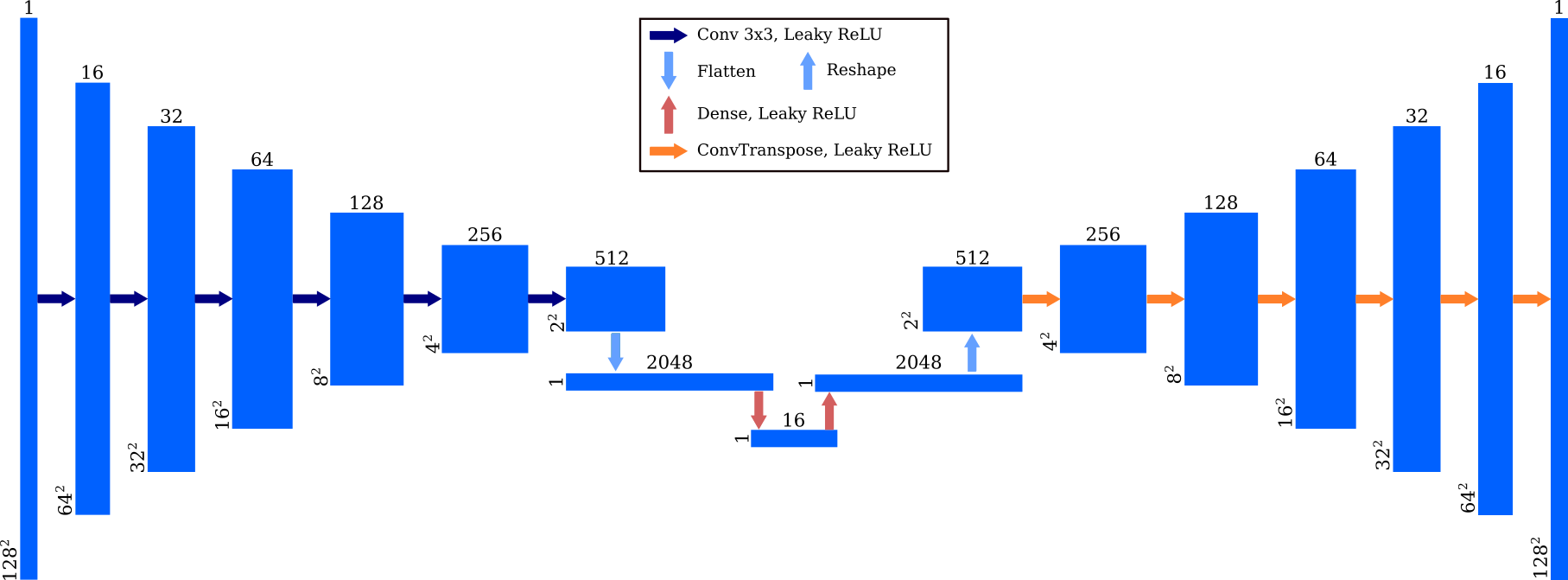}
\caption{\textbf{Autoencoder Network Architecture}. The anatomical autoencoder serves as a post-processing step and can be thought of as a separate encoder and decoder. }
\label{fig:anatae_model}
\end{figure}

This final neural network serves to adjust myocardial predictions such that they abide by anatomical guidelines. Expanding on the work in Painchaud~\etal~\cite{Painchaud2019}, we developed a binary function $\delta(\cdot)$ which uses different morphological operations to determine if a myocardium mask is anatomically correct. This function checks for convexity defects, holes in myocardium, circularity thresholds, number of objects, and myocardial wall thickness. The convolutional autoencoder is trained to reproduce myocardial segmentations after encoding them to a $d$-dimensional vector via a map $\phi$ (see fig.~\ref{fig:anatae_model}), which is approximately invertible: let the decoding function be $\phi^{-1}$. Given the limited data, the low-dimensional vectors of the masks populating the latent space of the autoencoder may not be sufficient to capture the diverse geometries of valid segmentations. Therefore, the latent space is augmented with a large number of $d$-dimensional vectors, $\bm{z}$, such that when decoded by the network, $\delta(\phi^{-1}(\bm{z})) = 1$.

\begin{figure}[ht]
\centering
\includegraphics[width=\textwidth]{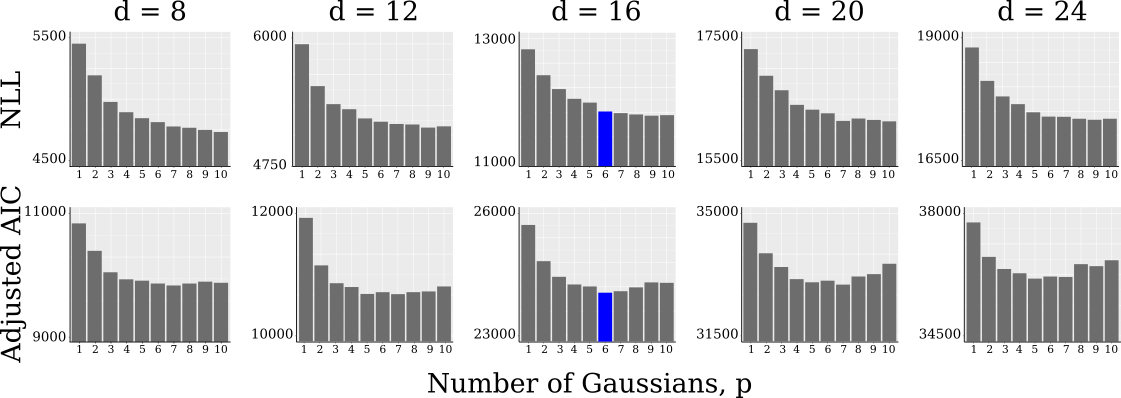}
\caption{\textbf{Autoencoder Latent Space Fit}. Ten different $p$-component (inset, x-axis) Gaussian mixture models are fit to the training data encoded by the autoencoder. The negative log-likelihood (NLL, top row) and Akaike information criterion scaled by the effective rank (Adjusted AIC, bottom row) are calculated for various latent space dimensions $d$ (columns) using 10-fold cross-validation. The dimension and number of components selected for the final model were $d=16$ and $p=6$ (blue) such that NLL plateaus, Adjusted AIC is low, and $d$ is high enough to ensure accurate reconstruction of the images by the autoencoder.}
\label{fig:gmm_modeling}
\end{figure}

We fit a Gaussian mixture model (GMM) with $k$ components to the training $d$-dimensional vectors in the latent space. We estimate $k=5$ and $d=16$ using the negative log likelihood (NLL) and adjusted Akaike information criterion (AIC)~\cite{Akaike1974} by cross-validation on the training set. In order to avoid penalizing high dimensional fits with many small singular values in the covariance matrix, we adjust the standard AIC by scaling the number of parameters by the effective rank $\Tr(\Sigma) / \sigma_{\text{max}}(\Sigma)$, where $\Tr$ is the trace, $\sigma_{\text{max}}$ is the spectral norm, and $\Sigma$ is the covariance matrix of a GMM component (fig.~\ref{fig:gmm_modeling}). When sampling from the new distribution, we reject a vector if, once decoded, the resulting mask does not pass the anatomical check $\delta$. Once trained, the autoencoder's latent space is populated using this re-sampling scheme with vectors which are ensured to decode to anatomically correct masks. When predicting, a new --- potentially incorrect --- mask, $\hat{I}$, such mask is first encoded (fig.~\ref{fig:network_arch}C) in the latent space to a vector $\hat{\bm{z}} = \phi(\hat{I}) \in \R^d$. If $\delta(\phi^{-1}(\hat{\bm{z}})) \neq 1$, that is, $\hat{I}$ does not encode and decode to an anatomically correct image, a nearest-neighbor algorithm is used to find the closest match $\hat{\bm{z}}_{NN}$ in the latent space $\hat{\bm{z}}_{NN} \coloneqq \argmin_{\bm{z} \text{ s.t. } \delta(\phi^{-1}(\bm{z}))=1} \norm{\bm{z} - \hat{\bm{z}}}^2$, where $\bm{z}$ varies over the constructed couples that satisfy anatomical constraints. Lastly, we define the final, anatomically correct segmentation as $\hat{I'} =\phi^{-1}(\hat{\bm{z}}^{*})$, where $\hat{\bm{z}}^{*} = \hat{\bm{z}} + \alpha^{*}(\hat{\bm{z}} - \hat{\bm{z}}_{NN})$, and $\alpha^{*}$ is the smallest $\alpha$ in $[0,1]$ such that $\delta[\phi^{-1}(\hat{\bm{z}} + \alpha(\hat{\bm{z}} - \hat{\bm{z}}_{NN})] = 1$ (fig.~\ref{fig:network_arch}D).

Our method for ensuring the anatomical accuracy of segmentations has key advantages over previous work. ACSNet uses a convolutional autoencoder with six $3\times3$ convolutional layers paired with leaky ReLU activation to encode to 16-D vectors; this architecture has significantly lower capacity than the ten-layer variational autoencoder implemented by Painchaud~\etal\cite{Painchaud2019} that derives a $\sigma$ and $\mu$ value for their 32-D latent distribution. Thus, ACSNet halves the number of latent dimensions used in Painchaud~\etal\cite{Painchaud2019}, demonstrating improved ability to represent images as smaller vectors. Further, one should be wary of the ability to generate more complex anatomical segmentations when re-sampling the space given the simple, unconstrained multi-variate normal assumption used in other work~\cite{Painchaud2019}. Besides anatomical consistency, our refined model of the distribution in encoded space allows, for example, for \qts{C}-shaped myocardium should these arise in basal slices. 

Different experts may select different vertical starting and stopping points when segmenting a patient's scan from apex to base. To standardize our diverse data, we apply three additional volumetric checks that account for the excess of images often included in a CMR scan, those below the apex or above the base of the LV. We compare ratios of myocardium and blood pool areas of each slice to identify the longest subsequence of relevant slices in the stack. Segmented volumes end at the index $i = \max(i_M,\ \min(i_C + 1, i_D))$. Here, $i_M$ refers to the final index in the largest subsequence of increasing slice area; $i_C$ represents the index of the first \qts{C}-shaped slice (a segmentation shape that can occur at the boundary of the ventricle and the atrium in the basal region); and $i_D$ represents the index of a large deviation in area between successive slices. This check allows incorporation of at most one \qts{C}-shaped slice and excludes slices above the base with no true region of interest, often predicted incorrectly as a small or excessively large region. The threshold for whether to include a slice in the subsequence, approximately a 60\% decrease in area, represents the 95th percentile of the difference in areas between all sequential slices. Final segmentations of patient scans now pass per-slice and per-volume anatomical constraints.

\subsection*{Training}
In total, $2,484$ 2-D images (LGE-CMR and \qts{LGE-like}) were used for training and $269$ LGE-CMR images were set aside for testing. The latter represented randomly selected LGE-CMR scans from $25\%$ of patient cohort.  Of note, there were no patients with multiple image modalities. No early stopping or other methods that learn from the validation set were used in training. All networks were trained on a NVIDIA Titan RTX GPU using Keras~\cite{keras} and Tensorflow~\cite{tensorflow}. Training input to the LV segmentation network included $1,360$ style-transferred ED cine-to-LGE images, as well as $1,124$ LGE-CMR images, reserving $25\%$ of the LGE-CMR scans for testing purposes. For the myocardium segmentation network, the training set used all patient scans with a corresponding enhancement segmentation, including style-transferred cine images with pseudo-scar. This subset amounted to $1,355$ \qts{LGE-like} ED cine images and $744$ of the LGE-CMR images for training the myocardium segmentation network, with the same validation set withheld as for the LV segmentation network. Finally, the autoencoder used ED cine and all LGE-CMR ground truth myocardium masks in the training set. 

Training of both the LV and the myocardium segmentation networks use the Adam optimizer~\cite{adam}. To prevent cine-derived LGE images from dominating the training set, we weigh these accordingly in the loss function. Thus, the loss function used is a weighted combination of the balanced cross-entropy loss and the Tversky loss~\cite{Salehi17}:
\begin{equation*}
\label{bce}
    l_1(p, \hat{p}) = -(\beta p\log(\hat{p}) + (1 - \beta)(1 - p)\log(1 - \hat{p}))
\end{equation*}
\begin{equation*}
\label{dice}
    l_2(p, \hat{p}) = 1 - \frac{2TP(p, \hat{p})}{2TP(p, \hat{p}) + \beta FP(p, \hat{p}) + (1-\beta) FN(p, \hat{p})},
\end{equation*}
where $p$ and $\hat{p}$ are pixel ground truth and predicted values, T/F P/N are true/false positive/negatives, and $\beta$ is a weight on the false positives. We avoid over-cropping after the first network by using a $\beta$ of $0.6$ for LV identification and account for two outputs in the myocardium segmentation network, placing higher weight on false positives, by using a $\beta$ of $0.4$. By optimizing over the mean of a per-pixel ($l_1$) and per-image ($l_2$) loss, each network is tailored towards the end goal of myocardial segmentation. 

\subsection*{Evaluation}
We evaluate standard imaging metrics to demonstrate the reliability of ACSNet across the entire scan. To do so, we calculate the Dice score of each slice's segmentation and use its location along the z-axis to place it into the first third, middle third, or upper third of the heart (corresponding to apex, mid-ventricle, and base). We demonstrate the clinical applicability of ACSNet segmentations through consistently high performance as opposed to a high average with poor-performing outliers.

Additionally, we evaluate LV volume calculation and quantification of scar using these segmentations. To calculate volume, nearest-neighbor interpolation is performed in the short-axis direction across each slice in a scan, accounting for pixel spacing in each slice. Volume includes myocardium and blood pool. When quantifying scar, we note that because ground truth enhancement masks include both dense core scar and heterogeneous gray zone, it can be assumed that the network picks up both areas of fibrotic tissue as well. We begin this analysis by cleaning up the enhanced myocardial predictions. The cleanup method was found by optimizing a function that applies morphological changes to segmentations predicted from the training set and compares the resulting Dice scores. We perform a single erosion with a $3\times3$ ellipsoid structuring element. Following the erosion, we keep up to five components in the segmentation, which each must be greater than a specified minimum area. After cleaning up the enhanced myocardial segmentations, we identify the dense core scar for quantification. Dense core scar is often calculated using a threshold of the full width at half maximum (FWHM) of myocardium intensity values in LGE-CMR images~\cite{SchmidtWu2007,JablonowskiWu2017}. By performing this threshold on both the enhancement predictions as well as the ground truth enhancement segmentations, we successfully split the gray zone from the core scar tissue. Finally, core scar tissue quantification is calculated as the summation of the number of pixels above the FWHM threshold. Results for both LV volume and scar volume are reported as relative MAE, which is the MAE divided by the respective volume.

\bibliography{main}

\section*{Acknowledgements} NT acknowledges support from NIH (grants R01HL142496, R01HL124893 and U01HL126273), a grant from the Leducq Foundation, and a grant for the Lowenstein Foundation. MM is grateful for support from the Simons Foundation, from NSF 2031985, 1837991, and AFOSR FA9550-20-1-0288. KCW is grateful for support from NIH grant HL103812 and HL132181.

\section*{Author contributions statement} DMP, HGA, MM and NAT designed the analysis. DMP and HGA developed the code and conducted the analysis. KCW provided annotated data and input on data interpretation, statistical analyses, and clinical aspects. All authors wrote the manuscript. 

\section*{Competing interests}

The authors have no competing interests to disclose.
\newpage

\begin{table}[ht]
\centering
\begin{tabular}{@{}lll@{}}
\toprule
\multicolumn{1}{c}{Method} & \multicolumn{1}{c}{MYO Dice Score} & \multicolumn{1}{c}{MYO Hausdorff Distance (mm)} \\ \midrule
ACSNet & \multicolumn{1}{r}{$0.79\pm0.02$} & \multicolumn{1}{r}{$6.70\pm0.53$} \\
Interobserver~\cite{Zhuang2016,Zhuang2019} & \multicolumn{1}{r}{0.76$\pm$0.08} & \multicolumn{1}{r}{12.50$\pm$5.38} \\
Zabihollahy~\etal\cite{Zabihollahy2020} & \multicolumn{1}{r}{0.85$\pm$0.03} & \multicolumn{1}{r}{19.21$\pm$4.74} \\
Yue~\etal\cite{Yue2019} & \multicolumn{1}{r}{0.76$\pm$0.23} & \multicolumn{1}{r}{11.04$\pm$5.82} \\
Roth~\etal\cite{Roth2020} & \multicolumn{1}{r}{0.78} & \multicolumn{1}{r}{16.30} \\
Mean Result of MS-CMRSeg MICCAI Challenge~\cite{Zhuang2020} & \multicolumn{1}{r}{0.77$\pm$0.10} & \multicolumn{1}{r}{18.06$\pm$12.18} \\
Chen, Ouyang\etal\cite{ChenOuyang2019} & \multicolumn{1}{r}{$0.83\pm0.04$} & \multicolumn{1}{r}{$12.45\pm3.14$} \\
\bottomrule
\end{tabular}
\caption{\label{tab:seg_comparison}Comparison of LGE-CMR segmentation results for the LV myocardium. All entries were rounded from the provided values to the nearest tenths place. Note: These sources use different datasets; Data for Interobserver~\cite{Zhuang2016,Zhuang2019}, Yue~\etal\cite{Yue2019}, Roth~\etal\cite{Roth2020}, and Chen, Ouyang~\etal\cite{ChenOuyang2019} are based on the 2019 CMRSeg MICCAI challenge~\cite{Zhuang2020} consisting of 2-D LGE-CMR \textit{and} corresponding steady-state free precision (bSSFP) from 45 patients, various subsets of whom were used as test sets. Zabihollahy~\etal\cite{Zabihollahy2020} used three orthogonal views of 34 subjects with 3-D LGE-CMR scans.}
\end{table}

\end{document}